\documentclass[10pt,
unsortedaddress,
twocolumn,
amsmath,
amssymb,
aps,
prl,
]{revtex4-2}
\usepackage{braket}
\usepackage{graphicx}
\usepackage{lipsum}
\usepackage{dcolumn}
\usepackage{bm,color}
\usepackage[normalem]{ulem}

\usepackage{amsfonts}

\usepackage{epigraph}

\usepackage{etoolbox}
\makeatletter
\newlength\epitextskip
\pretocmd{\@epitext}{\em}{}{}
\apptocmd{\@epitext}{\em}{}{}
\patchcmd{\epigraph}{\@epitext{#1}\\}{\@epitext{#1}\\[\epitextskip]}{}{}
\makeatother

\newcommand{\pms}{%
  \mathrel{%
    \vcenter{\offinterlineskip
      \ialign{##\cr$+$\cr\noalign{\kern-1.5pt}$-$\cr}%
    }%
  }%
}
\newcommand{\mps}{%
  \mathrel{%
    \vcenter{\offinterlineskip
      \ialign{##\cr$-$\cr\noalign{\kern -1.5pt}$+$\cr}%
    }%
  }%
}
\newcommand{\pmmp}{%
  \mathrel{%
    \vcenter{\offinterlineskip
      \ialign{##\cr$\pms$\cr\noalign{\kern -.5pt}$\mps$\cr}%
    }%
  }%
}
\newcommand{\pmpm}{%
  \mathrel{%
    \vcenter{\offinterlineskip
      \ialign{##\cr$\pms$\cr\noalign{\kern -.5pt}$\pms$\cr}%
    }%
  }%
}
\newcommand{\sam}[1]{\textcolor{black}{#1}}
\begin{document}

\title{Counterflow leads to roton spectra in locally interacting superfluids}

\author{Samuel Alperin}
\email{alperin@lanl.gov}
\author{Eddy Timmermans}\email{eddy@lanl.gov}

\affiliation{\vspace{1mm}  \mbox{Los Alamos National Laboratory, Los Alamos, New Mexico 87545, USA}}

\begin{abstract}
The dynamics of strongly interacting quantum fluids such as Helium II are fundamentally distinct from those of dilute, contact-interacting atomic Bose-Einstein condensates. Most dramatically, superfluids with finite-range interactions can support excitations with a roton-like dispersion, exhibiting a minimum at finite wavenumber. Here we introduce a mechanism through which roton spectra can emerge in superfluids without any nonlocal interactions, instead resulting from the collective excitations of two counterflowing, zero-range interacting condensates. As our mechanism relies only on the nonlinear dynamics of classical fields, this work opens the door to the realization of roton dynamics in the broad class of physical systems governed by coupled nonlinear-Schrödinger equations. 

\end{abstract}

\maketitle

Dispersion relations, which characterize the energy-momentum scaling of fundamental collective excitations, are a central tool in the study of many body systems.  In dilute atomic Bose-Einstein condensates (BECs), the structure of fundamental excitations is given by the Bogoliubov spectrum \cite{bogoliubov1947theory}, the structure of which depends on the lengthscale of underlying interatomic interactions. Aside from the recently realized dipolar BECs, atomic BECs generally have effectively zero-range (contact) interatomic interactions and are therefore well described in the mean field by the Gross-Pitaevskii equation (GPE); in this case the Bogoliubov spectrum reduces to its best known form, with linear dispersion at the origin indicating the collective excitation of phonons and a quadratic dispersion at higher momentum \cite{pethick2008Bose}. The monotonically increasing dispersion relation that follows from the absence of finite-range interactions stands in stark contrast to the dispersion relations associated with dense, strongly interacting quantum fluids such as superfluid Helium, which are characterized by the appearance of an energy minimum at finite momentum, the excitations around which are known as rotons \cite{landau41}.

The emergence of rotonic excitations significantly differentiates the phenomenology of finite-range-interacting superfluids from that of their simpler, contact-interacting cousins. For example, the ability of a superfluid to support roton excitations lowers the maximum superflow velocity \cite{landau41}, causes density oscillations to appear around defects \cite{staliunas2003transverse}, and most dramatically, can lead to the condensation of the rotons themselves \cite{Iordanskii1980Bose}. In the case of roton condensation, which is theorized to occur for suitably high roton density and low temperature, the superfluid adopts crystalline order while retaining global phase coherence \cite{melnikovsky2011bose}. Such a state, characterized by spontaneous breaking of both translational and U(1) symmetries, is known as a supersolid \cite{chester1970speculations,leggett1970can, boninsegni2012colloquium}.

While the search for evidence of a supersolid phase in Helium has remained inconclusive over many decades, \cite{boninsegni2012colloquium,prokof2005supersolid,saccani2012excitation}, advances in the creation and control of BECs of finite-range interacting lanthanide atoms has led to recent breakthroughs in the experimental study of roton dynamics and supersolidity: by controlling both the contact and finite-range dipolar inter-atomic interactions \cite{lu2011strongly,aikawa2012Bose,norcia2021developments}, the roton spectrum has been observed experimentally in BECs \cite{chomaz2018observation,natale2019excitation,petter2019probing}. Further, by tuning dipolar interactions past the point of roton instability, the critical density of roton excitations was achieved such that supersolid order was observed \cite{chomaz2019long,tanzi2019observation}.

Over the last three quarters of a century, from Landau's original theory to the recent realization of supersolidity, all works on rotons in \sam{uncharged superfluids} have, to our knowledge, presupposed the existence of underlying nonlocal interactions. In Feynman's seminal works on Helium, the roton dispersion was shown to follow directly from the form of the structure factor, which experiments had found to reveal finite-range density correlations \cite{feynman1956energy,feynman1955chapter,pines1989richard}. More recently, it was found that for weakly interacting Bosons, finite range interactions lead  generically to roton dispersions \cite{pomeau1993model,berloff1999motions}, a result exploited by the recent experiments on dilute dipolar gases \cite{chomaz2018observation,natale2019excitation,petter2019probing}. \sam{In the case of charged superfluids coupled an external gauge field, a roton-like dispersion can appear due to the appearance of an effective, anisotropic double-well potential; a number of works have exploited either laser coupling to dipole-dipole interactions or lattice shaking to achieve synthetic gauge field coupling and the resulting roton mode softening in neutral atomic BECs \cite{khamehchi2014measurement,ji2015softening,martone2012anisotropic,liao2015multicriticality,ravisankar2021effect,ha2015roton}. However, the ability to realize synthetic gauge-fields via dipole-dipole interactions is limited by the internal structure of the underlying atomic species, and the ability to realize such synthetic interactions via lattice shaking requires one to externally fix a particular spatial and temporal periodicity. Further, these techniques do not extend to more generic superfluid-like systems, such as exist in nonlinear optics.}

In this Letter, we introduce an entirely different path to the excitation of rotons, showing that they can emerge as the fundamental collective excitations of a mixture of two counterflowing, contact-interacting \sam{superfluids, without any need for long-range interactions or external gauge fields}. In the case that one superfluid is self-attractive and the other self-repulsive, a case in which for mutually stationary condensates is linearly unstable to collapse into quantum droplets \cite{petrov2015quantum}, we show that in 1D there is a critical relative velocity between condensate rest frames above which homogeneous densities become linearly stable and collective excitations become characterized by the emergence of a roton spectrum with finite energy gap. In the case where both condensates are self-repulsive, a roton minimum appears which always crosses into the regime of roton instability, raising the possibility of emergent supersolid like behaviours. While we discuss our results in the context of atomic BECs, they apply equally to any system described by coupled nonlinear Schrödinger equations, raising the possibility of realizing roton dynamics and even supersolid \sam{phases} in classical nonlinear optical settings.

\bigskip
\medskip
\noindent \textit{Two-Superfluid Dispersion --} We begin with the equations of motion for the homogeneous two-component atomic BEC in the mean field. Assuming only contact interactions, takes the form of the coupled GPE (or equivalently, the coupled nonlinear Schrödinger equations), which is written as

\begin{equation}
i \hbar \partial_t\psi_i = -\frac{\hbar^2}{2m_i}\nabla^2 \psi_i+ g_{ii}|\psi_i|^2 \psi_i+ g_{ij}|\psi_j|^2 \psi_i\label{eom}
\end{equation}
where $m_i$ are the atomic masses of the two species of Boson, and where $g_{ij}=2\pi \hbar^2 \left(\frac{1}{m_i}+\frac{1}{m_j}\right)a_{ij}$ is the interaction strength with scattering length $a_{ij}$. Physically, the two components represent either a mixture of two different atomic species or a single atomic species with two distinguishable internal states. In the latter case, the masses of the two components are equal, but the interaction strengths may in general be tuned independently \cite{mossman2024observation}. 

The solution for a homogeneous density and constant velocity field  takes the form $\psi_i\left(\boldsymbol{r},t\right)=\sqrt{n_i}e^{i\left( \boldsymbol{k}_i \cdot \boldsymbol{r} -\mu_i t/\hbar \right)}$, where $\mu_i$ is the chemical potential of condensate $i$ and $\boldsymbol{k}_i=m_i \boldsymbol{v}_i/\hbar$ is the de Broglie wavevector with flow velocity $\boldsymbol{v}_i$. The chemical potential can then be written, with \sam{$i \neq j$}, as 

\begin{equation}
    \mu_i=\frac{\hbar^2 \boldsymbol{k}\sam{_i}^2}{2m_i}+g_{ii} n_i+g_{ij} n_j.
    \label{chem}
\end{equation}
Writing the perturbed homogeneous state as $\sam{\delta}\psi_i=\left(\sqrt{n_i}+\delta_i\right)e^{i\left( \boldsymbol{k}_i \cdot \boldsymbol{r} -\mu_i t/\hbar \right)}$ where $|\delta_i|\ll |\psi_i|$, we substitute into Eq. \ref{eom}. Linearizing and using Eq. \ref{chem} yields the perturbation equation

\begin{equation}
\begin{split}
    i\hbar \frac{D_i}{Dt} \delta\psi_i=-\frac{\hbar^2}{2m_i}\nabla&^2\delta\psi_i+g_{ii}n_i\left(\delta\psi_i+\delta\psi_i^*\right)\\&+g_{ij}\sqrt{n_in_j}\left(\delta\psi_j+\delta\psi_j^*\right)
    \end{split}
\end{equation}
where $\frac{D_i}{Dt}=\left(\partial_t+\boldsymbol{v}_i\cdot \nabla\right)$ is the convective derivative. For simplicity, we introduce the the scaled sum and difference of the complex linearized field amplitudes  $\psi_i$ and its complex conjugate

\begin{eqnarray}
    \delta X_i = \frac{\delta\psi_i+\delta\psi_i^*}{\sqrt{2}}\\\delta P_i = \frac{\delta\psi_i-\delta\psi_i^*}{i\sqrt{2}}
\end{eqnarray}
which correspond to the real-valued fields describing the deviations in density and phase, respectively. After some rearranging, this allows us to rewrite the linearized equations of motion as
\begin{align}
    \frac{D_i^2}{Dt^2}\delta X_i&=\frac{\hbar\nabla^2}{2m_i}\left(-\frac{\hbar\nabla^2}{2m_i}+\frac{2g_{ii}n_i}{\hbar}\right)\delta X_i \nonumber \\ &-\left(\frac{\hbar\nabla^2}{2m_i}\right)\left(\ \frac{\sqrt{n_in_j}}{\hbar}\right) \delta X_j
\end{align}

As these field amplitudes are real-valued, we can take the standing wave ansatz by substituting $\delta X_i=x_i \cos(\boldsymbol{k}\cdot\boldsymbol{r}-\omega t)$ and $\delta P_i=p_i \sin(\boldsymbol{k}\cdot\boldsymbol{r}-\omega t$), and solve for the dispersion relation, which takes the general form

\begin{align}
    g c^2_1 c^2_2 k^4=\prod_{i=1}^{2}\left(( \omega+\boldsymbol{v}_i \cdot \boldsymbol{k})^2-\omega^2_{B,i}   \right)
    \label{disp1}
\end{align}
in which we denote the single condensate Bogoliubov frequencies as $\omega^2_{B,i}=k^2 c^2_i(1+\xi^2_i k^2)$, where the speed of sound and the healing length of condensate $i$ are defined as $c_i=\sqrt{g_{ii}n_i/m_i}$ and $\xi_i=\hbar/(2m_ic_i)$. We define the  interaction parameter $g=\frac{g_{12}^2}{g_{11} g_{22}}$, and note that the sign of $g$ depends only on whether or not $g_{11}$ and $g_{22}$ have the same sign.

While the dispersion relation defined by Eq. \ref{disp1} can be written analytically  as a function $\omega(k)$ without further assumptions \footnote{Unfortunately the generic dispersion relation, though analytically solvable, is far too long to print.}, for conceptual clarity we will from this point consider the simplest case, in which the two condensates components have the same mass and number density, so that $\omega_{B,1}=\omega_{B,2}=\omega_B$ and $c_1=c_2=c$. Physically, this is the case where the two condensates wavefunctions represent different internal states of the same atomic species, with interactions and velocities tuned independently, as has been achieved in past experiments \cite{mossman2024observation}. Denoting the counterflow velocity as $v=v_2-v_1$ and restricting ourselves to one spatial dimension \footnote{\sam{Note that while Bose-Einstein condensation is forbidden in 1D by the Mermin-Wagner theorem, the 1D-Gross-Pitaevskii equation remains a highly descriptive model of atomic BECs confined to the quasi-1D regime \cite{becker2008oscillations,stellmer2008collisions,strecker2002formation}. This model also applies directly to nonlinear optical propagation in two-mode fiber, a system which is naturally 1D.}}, we can write the dispersion succinctly as the four branch solution \footnote{The four branches of the solution are given by the four combinations of signs inside and outide the outermost root.}

\begin{equation}
   \omega(k)=   \pmpm   \sqrt{\pmmp\sqrt{g c^4 k^4+4 k^2 v^2 \omega_B^2}+k^2 v^2+\omega_B^2}.
   \label{disp}
\end{equation}

The main results of this Letter will follow directly from the characterization and analysis of Eq. \ref{disp}. As will be seen, the physics of this system depends significantly on the sign of the interaction parameter $g$; we will begin with the analyses of the case in which $g<0$, corresponding to the mixture of self-attractive and self-repulsive condensates, which is well known to be unstable to collapse when mutually at rest.

\begin{figure}[h!]
\centering
\includegraphics[width=\columnwidth]{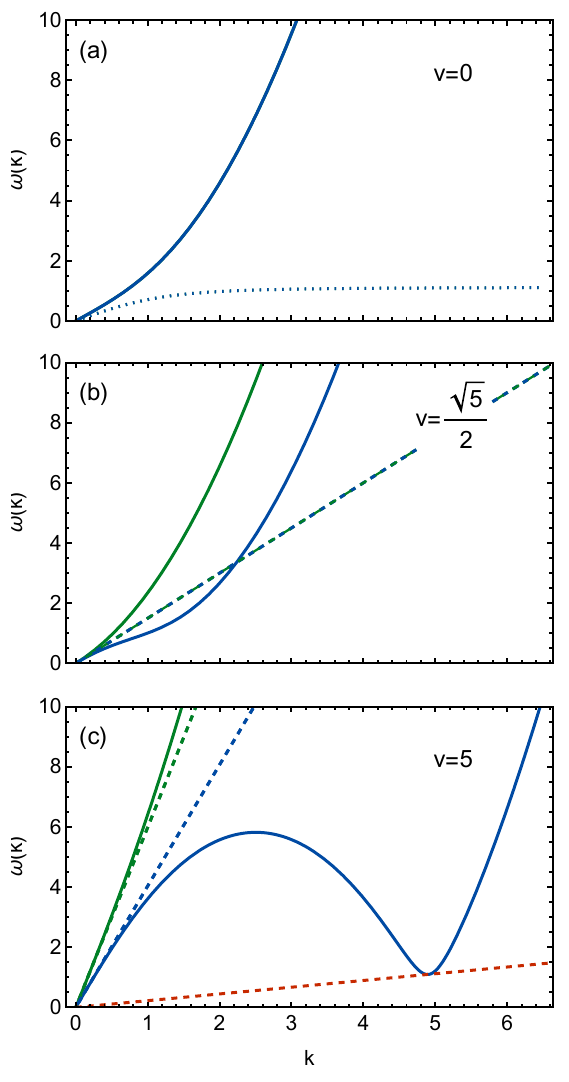}
 \caption{Dispersion relation characterizing the fundamental excitations of counterflowing superfluids, given by Eq. \ref{disp}, where one superfluid component has positive scattering length, and the other has negative scattering length ($g<0$). In the stationary case $v=0$ (a), there are unbounded positive imaginary frequencies (dotted) which signify instability, as is expected when $g<0$. (b) However, for counterflow velocity at the threshold $v_{cc}$, the system becomes dynamically stable, and despite the appearance of two distinct dispersion branches, a single real-valued phonon velocity appears (dashed). For velocities above the threshold (c), all frequencies are real, signifying linear stability, and we see two distinct phonon modes (dashed), in addition to an emergent roton mode. Here  $c=\xi=1$.}
 \label{disp_fig}
\end{figure}

\bigskip
\noindent \textit{Gapped Roton Regime --} 
Linear instability exists when frequencies $\omega(k)$ given by  Eq. \ref{disp} are complex with positive imaginary component, for some wavenumbers. When $g<0$, it follows directly from  Eq. \ref{disp} that there exists a critical velocity $v_c=\frac{c\sqrt{|g|}}{2}$ such that $\omega(k) \in \mathbb{R}$ for $v\geq v_c$l. As can be seen in Fig. \ref{disp_fig}, for $v<v_c$ frequencies have positive imaginary components (Fig. \ref{disp_fig}.a), signaling linear instability; for $v=v_c$ the dispersion is real-valued and thus stable for all wavenumbers. While there are two excitation branches, both modes share the same slope at the origin, indicating a single effective phonon mode (Fig. \ref{disp_fig}.b). Above the threshold, three quasiparticles emerge, with a roton minimum appearing in addition to two distinct phonon modes. 

The phonon modes are fully characterized by their sound velocities, and the roton by the location of the minimum. We start by calculating the phonon velocities, which are defined as the rate of change of the energy as a function of wavenumber at zero energy and wavenumber. These have the form

\begin{align}
v_{ph}&=\lim_{k \rightarrow 0^+}\bigl( \partial_k\omega(k) \bigl) \\
&=\frac{\pm  g+c^2 \left(\sqrt{4 c^2 v^2+c^4 g}\pm 4 v^2\right)+v^2 \sqrt{4 c^2 v^2+c^4 g}}{\sqrt{4 c^2 v^2+c^4 g} \sqrt{\pm  \sqrt{4 c^2 v^2+c^4 g}+c^2+v^2}},
\end{align}
the structure of which is nontrivial, having three distinct forms of solutions, shown in Fig. \ref{disp_fig}: (a)
below the critical velocity, the phonon velocities are complex; (b) at the critical velocity the phonon velocity is real and single-valued; and (c) above the critical velocity there emerge two distinct, real phonon velocities.

Noting that $\sqrt{1+k_{rot}^2 \xi ^2}\approx\ k_{rot} \xi$ for physical parameters, we can well approximate the location of the roton minimum as

\begin{align}
&k_{rot}=\sqrt{\frac{\sqrt{c^4 g+v^4}}{2c^2 \xi ^2}+\frac{v^2}{2 c^2 \xi ^2}}\label{krot} \\ \label{wrot}
   & \omega(k_{rot})=  \frac{1}{2c\xi^2}\biggl(g c^4 \xi ^2+2 c^2 \left(\beta +\xi ^2 v^2\right)+4 \xi ^2 v^4+\\ &\nonumber4 \beta  v^2- 2 \xi  \biggl(g^2  \xi ^2 c^8+4 g  \xi ^2 v^2c^6 +4g v^2 c^4   \left(\beta +2 \xi ^2 v^2\right)+\\ \nonumber&8 c^2 v^4 \left(\beta +\xi ^2 v^2\right)+8 v^6 \left(\beta +\xi ^2 v^2\right)\biggl)^{1/2}\biggl)^{1/2}
\end{align}
where $\beta=\xi ^2\sqrt{g c^4+v^4}$.
From simple arguments, Landau showed that it is the ratio of the roton energy and wavenumber, a velocity known as the Landau critical velocity, below which superfluid Helium can flow in a channel without dissipation \cite{landau41,mcclintock1995landau}. Here this velocity can be written by substituting Eqs. \ref{krot}-\ref{wrot} into its definition $\sam{v_{lc}}=\omega(k_{rot})/k_{rot}$. This estimate of the Landau critical velocity is shown by the slope of the red dashed line in Fig. \ref{disp_fig}.c: any excitation with higher-than-critical velocity crosses into the regime in which it is energetically favorable for rotons to form.

\begin{figure}[h!]
\centering
\includegraphics[width=\columnwidth]{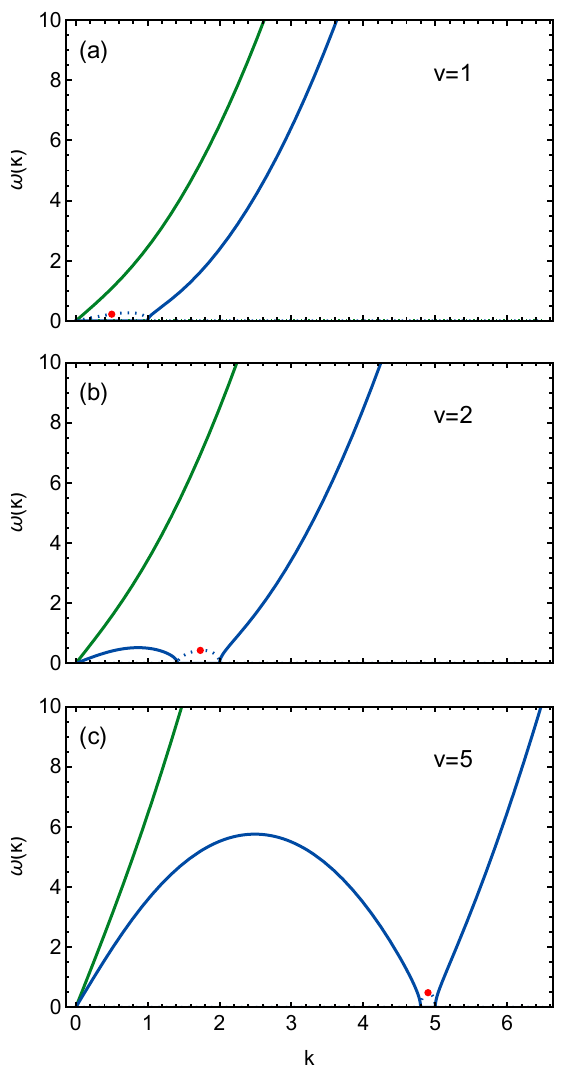}
\caption{Dispersion relation for two counterflowing, contact interacting superfluids, where both fluids are self-repulsive ($g>0$) and miscible at $v=0$. For counterflow above an initial instability threshold (a), a region of wavenumberss $0\leq k \leq k_{u2}$ become linearly unstable (dotted). However, above a second threshold velocity (b), the instability region become bounded on both sides by finite wavenumbers $0 < k_{u1} \leq k \leq k_{u2}$, corresponding to the appearance of a roton instability. For increasing counterflow velocity (c), the range of unstable wavenumbers shrinks, tending asymptotically to $k_{u2}-k_{u1}=0$. Red points mark the imaginary value of the frequency at the midpoint of unstable wavenumbers. Note that the roton wavenumber in (c) is identical to that of the system in Fig. \ref{disp_fig} (c), which differs only in the sign of $g$. Here  $c=\xi=1$.}
 \label{fig2}
\end{figure}

\bigskip
\noindent \textit{Roton Instability Regime --} 
We now consider the case where $g>0$, which corresponds to the case where both components of the superfluid have self-interactions with the same sign. For a miscible mixture of two superfluids, it was shown by Abad et al. that linear stability is lost when there is a relative velocity between the components greater than a critical value \cite{abad2015counter}.  Indeed, we find that above a critical counterflow velocity, the system becomes unstable for a finite range of wavenumbers  centered at $k=0$ (see Fig. \ref{fig2}.A). However, going beyond the long wavelength analysis of Abad et al, we find that as the counterflow velocity surpasses a second critical velocity the range of unstable wavenumbers ($k_{u1}\leq k \leq k_{u2}$) stops including $k=0$; as the velocity increases, the width of k-space that is unstable, $\Delta k_{u}=k_2-k_1$, tends to zero asymptotically as 
\begin{equation}
    \Delta k_{u}=\frac{c \sqrt{g}}{\xi  v},
\end{equation}
with mean unstable wavenumber increasing asymptotically as 
\begin{equation}
    \langle k_{u}\rangle=\frac{v}{c \xi}.
\end{equation}
The Lyapunov exponent, which characterizes the timescale at which instability sets in, is defined as $\lambda = \max\left[\mathfrak{I}(\omega(k)\right]$. Noting that the wavenumber corresponding to the largest imaginary frequency quickly converges to $ \langle k_{u}\rangle$ with increasing velocity, we can substitute the mean wavenumber into the dispersion and take the large velocity limit, yielding the limit of the Lyapunov exponent
\begin{equation}
    \lim_{v\to\infty}\lambda(v)= \max_{v}\lambda(v) =\frac{c \sqrt{g}}{2 \xi }.
\end{equation}

These asymptotic characteristics apply to both miscible and immiscible mixtures of superfluids: for suitable counterflow velocity, the instability, existing at an increasingly narrow range of lengthscales as $v \rightarrow \infty$, takes the form of the  roton instability, where the roton energy gap crosses into the complex plane. This corresponds to the scenario in which it costs no energy to excite rotons. Such an instability was exploited in recent dipolar condensate experiments, to generate the roton density necessary for the condensation of the fundamental collective excitations \cite{bottcher2019transient}.
In the case of the immiscible mixture, the roton instability represents a direct extension of the phase separation instability, a well established pattern forming instability that can lead to the formation of droplets in stationary mixtures of BECs \cite{timmermans1998phase}:  as the counterflow velocity is increased from zero in the immiscible mixture, the structure of the phase separation instability is continuously transformed into that of the roton instability.

Throughout the decades of work on rotons, the existence of these fundamental collective excitations has always followed from the nonlocal nature of the underlying interactions. In contrast, we have shown that roton dispersions are significantly more generic, characterizing the fundamental excitations of counterflowing, locally interacting nonlinear fields. While the fundamental nature of rotonic excitations is of significant inherent interest, our work also has implications in the study of supersolidity, with the existence of a tunable roton minimum suggesting the possibility of roton condensation, and as a result supersolidity, not only in strictly contact-interacting BECs, but also in fully classical systems that obey two-component nonlinear Schr\"odinger dynamics, such as coupled nonlinear optical waveguides. We leave the explicit construction and analysis of such classical, nonlinear dynamical supersolid states to future work.

\medskip
\noindent 

S.A. would like to thank the Isaac Newton Institute for Mathematical Sciences, Cambridge, for support and hospitality during the programme ``Emergent phenomena in nonlinear dispersive waves", where work on this paper was undertaken. This work was supported by EPSRC grant EP/R014604/1, and by Los Alamos National Laboratory LDRD program grant 20230865PRD3.

\end{document}